\documentstyle[aps,prl,twocolumn]{revtex}
\draft

\begin{document}
\title{Can Neutrinos be Majorana Particles?}
\author{Steen Hannestad}
\address{Theoretical Astrophysics Center and
Institute of Physics and Astronomy,
University of Aarhus, 
DK-8000 \AA rhus C, Denmark}
\date{\today}
\maketitle

\begin{abstract}
It was recently claimed that the observed physical neutrinos
cannot be of Majorana type because such particles lack
vector interactions. Contrary to this claim we show that if
the Majorana neutrino is massless it is indistinguishable from the
Dirac neutrino so that observed physical neutrinos may 
equally well be of Majorana or Dirac type.
\end{abstract}


Neutrinos are the only known fermions without
charge and therefore they may in principle be their own antiparticles.
If this is the case, only two free parameters are needed to 
describe the neutrino, a left handed and a right handed component.
This is the well known Majorana type neutrino
\cite{mohapatra}. The standard fermion
description of a neutrino is the Dirac neutrino, consisting of
four degrees of freedom, left and right handed particles and
antiparticles \cite{mohapatra}.
However, two of these components are sterile because only the
left handed neutrino and the right handed antineutrino interact
if the neutrino is massless. Even though both Dirac and Majorana
neutrinos are described by only two different parameters there
is the fundamental difference that the Majorana neutrino field
is invariant under $CPT$ \cite{mohapatra}. 
Therefore one might suspect that the
Dirac and Majorana descriptions are not equivalent even for
zero neutrino mass.

Indeed, it was recently claimed that the observed physical neutrinos
of the weak interaction cannot be Majorana particles because
physical Majorana neutrinos should be unpolarized \cite{plaga}.
Furthermore such particles do not possess vector neutral current
interactions.
Therefore they should have interactions that are quite different
from Dirac neutrinos and be excluded by the CHARM II 
measurement of the neutrino neutral current coupling \cite{charm}.

The argument is that the Majorana field is a linear combination
of left handed and right handed parts
\begin{equation}
\psi_{M} = \frac{1}{\sqrt{2}} (\psi_{L} + \psi_{L}^{CPT})
= \frac{1}{\sqrt{2}} (\psi_{L}^{M} + \psi_{R}^{M}),
\label{field}
\end{equation}
where $\psi_{L}$ is a left handed neutrino field.

In Ref.\ \cite{plaga} it is claimed that the same should be true
for the Majorana states.
However, Eq.\ (\ref{field}) is not equivalent to saying that the physical 
Majorana states are always unpolarized. 
Indeed, the Majorana field should be invariant under the $CPT$
operation. However, the physical Majorana state reverses its
spin under $CPT$ \cite{mohapatra}
\begin{equation}
\Theta \mid \nu (p,s)> = \eta_{\Theta} \mid \nu (p,-s)>,
\end{equation}
where $\Theta = CPT$ and $\eta_{\Theta}$ is a phase factor.
Therefore, contrary to what the author of
Ref.\ \cite{plaga} claims, the states do not have to be invariant
under $CPT$ and therefore physical Majorana neutrinos are not
required to be unpolarized. This just shows how important it is
to distinguish between the field and the states. The field is by
definition an operator, defined by the creation and annihilation
operators, whereas the states are defined by the action of the
creation operator on vacuum \cite{kim}

Now, let us consider the situation with the CHARM II experiment.
This experiment uses muon neutrino and antineutrino
scattering on target electrons to measure the neutral weak
current coupling constants, $A$ and $V$.
The neutrinos are produced
via the charged weak current decays of charged kaons and pions \cite{charm}
and therefore they are born with definite helicity. 
This is simply because of the $V-A$ structure of the charged
weak current
\begin{equation}
\newfont{\sss}{cmsy10}
\text{\begin{sss} L \end{sss}} = \overline{e} \gamma_{\mu}
(1 - \gamma_{5}) \nu_{M}.
\end{equation}
Negatively charged kaons and pions have to produce muons and
right handed muon neutrinos
 because the right handed current is not active
in weak interactions, whereas positively charged kaons and pions
necessarily produce lefthanded muon neutrinos \cite{grotz}.
The calculation of the cross section for neutral current scattering
of a fully polarized Majorana neutrino on an electron then gives,
$\nu_{\mu} e^{-} \rightarrow \nu_{\mu} e^{-}$, is
straightforward and gives
\begin{eqnarray}
\mid \! M \! \mid^{2} & = & 4 G_{F}^{2}
      [(C_{A}-C_{V})^{2} (p_{\nu_{\mu,1}} \cdot p_{e,2})^{2} \\ \nonumber 
& &    +(C_{A}+C_{V})^{2} (p_{\nu_{\mu,1}} \cdot p_{e,1})^{2}],
\end{eqnarray}
where index 1 indicates incoming particles and index 2 outgoing particles.
This exactly equivalent to the corresponding Dirac matrix
element \cite{hannestad}.
This means that the vector interaction is reintroduced
by the definite initial chirality of the Majorana neutrino even though
one may show that the vector part of the neutral
current disappears by using $CPT$ conservation arguments \cite{shrock}. 
So even if
they are detected via the neutral current scattering on electrons,
the Majorana neutrinos will be completely indistinguishable
from Dirac neutrinos.

If the Majorana neutrino was indeed unpolarized \cite{early}, the
corresponding matrixelement would be
\begin{eqnarray}
\mid \! M \! \mid^{2} & = & 4 G_{F}^{2}
      [(C_{A}+C_{V})^{2} \{(p_{\nu_{\mu,1}} \cdot p_{e,2})^{2} \\ 
& &       + (p_{\nu_{\mu,1}} \cdot p_{e,1})^{2}\}], \nonumber
\end{eqnarray}
leading to a completely different cross section.

Thus, the Majorana neutrinos produced by charged currents are
 chiral eigenstates 
and therefore fully polarized because
of the lefthanded nature of the charged weak interaction, even though
the Majorana field is a linear combination of left and right handed parts.
The claim that the observed neutrinos of the weak interaction
cannot be Majorana particles is therefore not correct.

For massive neutrinos the situation is not so simple because there exist
no chiral eigenstates. This leads to detectable differences between
Dirac and Majorana neutrinos, because there are 4 physical components
for a massive Dirac neutrino \cite{gaemers}
\footnote{One major difference between massive Majorana and Dirac 
neutrinos is that helicity flipping 
involves violation of lepton number for Majorana neutrinos, whereas it
does not for Dirac neutrinos.}.
In this case one should use a polarization density matrix formalism
\cite{garavaglia} in order to do the calculations.

\acknowledgements
I thank Georg Raffelt and Jes Madsen for enlightening discussions on this
subject. Financial support by the Theoretical Astrophysics Center
under the Danish National Research Foundation is recognised.


\begin{references}

\bibitem{mohapatra}See for example R. N. Mohapatra and P. B. Pal,
{\it Massive Neutrinos in Physics and Astrophysics},
World Scientific (1991); C. W. Kim and A. Pevsner,
{\it Neutrinos in Physics and Astrophysics},
Harwood Academic Publishers (1993).
\bibitem{plaga}R. Plaga, Report No. hep-ph/9610545, 1996.
\bibitem{charm}P. Vilain {\it et al.}, Phys.\ Lett.\ {\bf 335B},
246 (1994).
\bibitem{kim}For a discussion of this see for example the monograph
by C. W. Kim and A. Pevsner cited above
\bibitem{grotz}K. Grotz and H. V. Klapdor, {\it The Weak 
Interaction in Nuclear, Particle and Astrophysics},
Adam Hilger (1990).
\bibitem{hannestad}S. Hannestad and J. Madsen, 
Phys.\ Rev.\ D\ {\bf 52} 1764 (1995).
\bibitem{shrock}R. E. Shrock, Phys.\ Rev.\ D\ {\bf 28}, 1023 (1984).
\bibitem{early}One case where Majorana neutrinos may be regarded
as unpolarized is
the early Universe, where the left and righthanded
components are equally abundant and may be treated as a single
unpolarized species. See for example B. D. Fields, K. Kainulainen,
and K. A. Olive, Report No. hep-ph/9512321, 1995.
\bibitem{gaemers}See for example K. J. F. Gaemers, R. Gandhi, and
J. M. Lattimer, Phys.\ Rev.\ D\ {\bf 40}, 309 (1989);
K. Grotz and H. V. Klapdor, {\it The Weak Interaction in Nuclear,
Particle and Astrophysics}, Adam Hilger 1990.
\bibitem{garavaglia}T. Garavaglia, Phys.\ Rev.\ D\ {\bf 29}, 387 (1984).
\end{references}
\end{document}